\documentclass[10pt,a4paper,oneside]{article}
\usepackage[cp850]{inputenc}
\usepackage{amsmath,amsfonts,latexsym,amssymb}
\usepackage{epic,eepic}
\usepackage{epsfig}
\usepackage{graphicx}

\DeclareMathOperator{\str}{str}

\DeclareMathOperator{\diag}{diag}

\DeclareMathOperator*{\Simi}{\sim}

\newcommand{\vect}[1]{{\mathbf #1}}
\newcommand{\vectgr}[1]{{\boldsymbol#1}}    %Greek letter in bold

\newcommand{\Frac}[2]{\displaystyle\frac{#1}{#2}}
\newcommand{\smc}[1]{\text{\sc{#1}}}

\title{\vspace{-3\baselineskip}%
       Tail States in Disordered Superconductors with Magnetic
       Impurities: the Unitarity Limit}

\author{\small F. M. Marchetti${}^{1,2}$ and B. D. Simons${}^{1}$ \\
       \small ${}^{1}$ \emph{Cavendish Laboratory, Madingley Road,
       Cambridge CB3 \ OHE, UK }\\
       \small ${}^{2}$ \emph{Scuola Normale Superiore, Piazza dei
       Cavalieri 7, 56126 Pisa, Italy}}

\date{\small November 9, 2001}

\begin{document}

\maketitle

\begin{abstract}
  \emph{When subject to a weak magnetic impurity distribution, the 
  order parameter and quasi-particle energy gap of a weakly disordered 
  bulk $s$-wave superconductor are suppressed. In the Born scattering 
  limit, recent investigations have shown that `optimal fluctuations'
  of the random impurity potential can lead to the nucleation of 
  `domains' of localised states within the gap region predicted by the
  conventional Abrikosov-Gor'kov mean-field theory, rendering the
  superconducting system gapless at any finite impurity
  concentration. By implementing a field theoretic scheme tailored to
  the weakly disordered system, the aim of the present paper is to
  extend this analysis to the consideration of magnetic impurities in
  the unitarity scattering limit. This investigation reveals that the
  qualitative behaviour is maintained while the density of states
  exhibits a rich structure.}
\end{abstract}

\section{Introduction}
\label{sec:3intr}
In the absence of Coulomb interaction effects, the spectral and 
transport properties of a bulk singlet $s$-wave superconductor are 
largely insensitive to the presence of a weak non-magnetic impurity
potential. This effect, which is ascribed to the Anderson 
theorem~\cite{anderson1}, limits the influence of long-range phase 
coherence phenomena to situations in which low-energy quasi-particles 
persist: notably, the physics of hybrid SN-compounds, and those which 
exhibit unconventional (e.g. $d$-wave) symmetry. However, another 
method of inducing low-energy quasi-particle states in the disordered
superconducting environment is to impose an external time-reversal
symmetry breaking perturbation which has a pair-breaking effect on the
condensate.

Considering mechanisms of time-reversal symmetry breaking, it is possible
to conceive of at least two distinct physical situations. The first is the
imposition of a homogeneous magnetic field: the diamagnetic properties
of the superconductor limit considerations to either a vortex phase of
a type II superconductor, or to superconductors whose lateral dimension is 
smaller than the penetration depth. In each case, field lines are able to 
penetrate the sample. A second method of breaking the intrinsic time-reversal 
symmetry of the system is to impose a magnetic impurity distribution. With
both mechanisms, the perturbation acts with opposite sign on the two members 
of the Cooper pair. In the first case, the paramagnetic term in the single 
particle Hamiltonian reverses sign under $\vect{p}\to -\vect{p}$. In the 
second case, the spin of the magnetic impurity acts on different spin
components with different sign. Building on the existing literature, the 
aim of this paper is to explore the quasi-particle properties of a weakly 
disordered superconductor subject to a magnetic impurity distribution with 
unitarity limit scattering.

\subsection{Background: Abrikosov-Gor'kov Theory}
\label{sec:bscag}
In the earliest work in this area, attention was focussed on the influence 
of a weak magnetic impurity distribution in which the influence of disorder
could be treated in the Born approximation. In a seminal work by Abrikosov 
and Gor'kov~\cite{abrikosov_gorkov}, it was shown that the pair-breaking
potential brings about only a gradual suppression of the superconducting 
order parameter. More surprisingly, according to the self-consistent 
mean-field theory, the energy gap in the quasi-particle density of states
(DoS) is suppressed more rapidly than the order parameter, admitting a region 
in the phase diagram where the superconductor exhibits a `gapless' phase. 
More precisely, defining the dimensionless control parameter,
\begin{equation*}
  \zeta = \Frac{1}{\tau_s |\Delta|} \; ,
\end{equation*}
where $|\Delta|$ represents the self-consistent bulk order parameter,
and $1/\tau_s$ denotes the Born scattering rate due to the magnetic
impurities, the Abrikosov-Gor'kov mean-field theory shows the energy
gap to vary as $\epsilon_g =|\Delta|(1-\zeta^{2/3})^{3/2}$,
showing an onset of the gapless region when $\zeta=1$. Soon after its 
introduction, it was realised that the general Abrikosov-Gor'kov scheme 
applies equally to other mechanisms of pair-breaking (such as that imposed 
by a uniform magnetic field in a thin film or by a supercurrent) --- 
requiring only a reinterpretation of the dimensionless parameter 
$\zeta$ (see, e.g., Ref.~\cite{tinkham}).

In later works~\cite{yu,shiba,rusinov,salkola_balatsky,flatte_byers},
various authors explored the influence of isolated magnetic
impurities. In particular, in the unitarity limit, it was shown that a
single \emph{classical} magnetic impurity of spin $S$ leads to the
local suppression of the order
parameter~\cite{salkola_balatsky,flatte_byers} and nucleates a bound
sub-gap quasi-particle state at energy~\cite{shiba}
\begin{displaymath}
  \Frac{\epsilon_{\smc{b}}}{|\Delta|} = \Frac{|1 - \alpha|}{1 +
  \alpha} \; ,
\end{displaymath}
where, defining the density of states $\nu=1/(L^d \delta)$ of a normal
conductor, with $\delta$ being the single-particle level spacing,
$\alpha = (\pi \nu J L^d |\vect{S}|)^2$ represents the dimensionless
scattering amplitude associated with the magnetic impurity.

\begin{figure}
\begin{center}
\includegraphics[width=1\linewidth,angle=0]{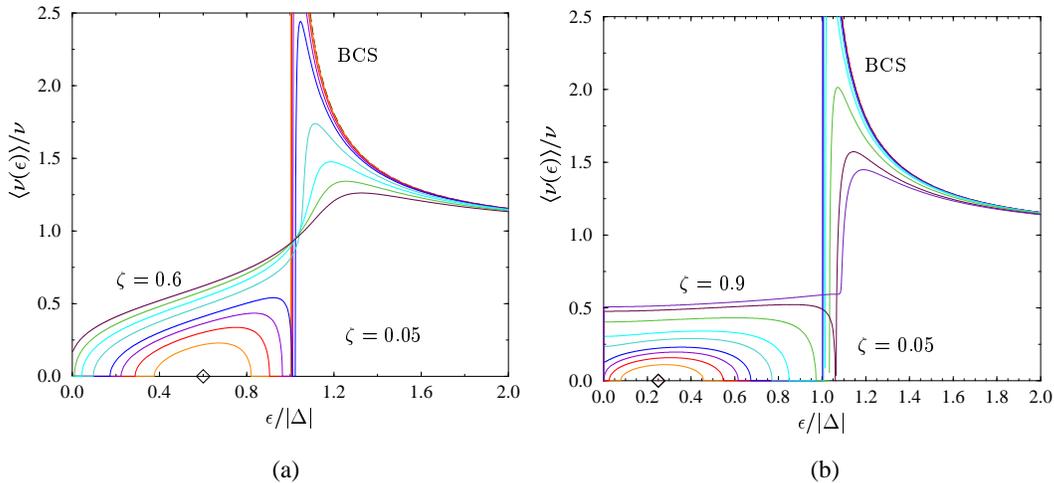}
\end{center}
\caption{\small 
        Quasi-particle density of states of a disordered superconductor 
        with magnetic impurities drawn from a Poissonian distribution;
        (a) $\alpha=0.25$ 
        ($\zeta_1 < \zeta_0$), the value of $\zeta$ is increased
        from $\zeta = 0.05$ to $\zeta = 0.6$. The localised excited
        state for the one impurity problem is located at
        $\epsilon_{\smc{b}}/|\Delta| = 0.6$.
        (b) $\alpha=0.6$ ($\zeta_0 < \zeta_1$), the value of
        $\zeta$ is increased from $\zeta = 0.05$ to $\zeta = 0.9$. The
        localised excited state for the one impurity problem is located at
        $\epsilon_{\smc{b}}/|\Delta| = 0.25$.}
\label{fig:dos1u}
\end{figure}

For a finite impurity concentration, the sub-gap states weakly
overlap, hybridise, and broaden into a band~\cite{shiba} centered on
energy $\epsilon_{\smc{b}}$. Here, as in the Abrikosov-Gor'kov
theory, the mean-field theory again predicts a gradual suppression of
the quasi-classical energy gap $\epsilon_g$, with the superconductor
entering the gapless phase when $\zeta = \zeta_0 \equiv
(1-\alpha)^2$. Increasing the magnetic impurity concentration
$\alpha$, two qualitatively different situations can be realised: in
the first case, when $\zeta$ reaches the value $\zeta_1$, the impurity
band can merge with the continuum of bulk states before the system
enters the gapless phase ($\zeta_1 < \zeta_0$), while, in the second
case ($\zeta_0 < \zeta_1$), the opposite situation pertains (see
Fig.~\ref{fig:dos1u}).

\subsection{Beyond Mean-Field Theory}
\label{sec:beymf}
Despite the success of the Abrikosov-Gor'kov mean-field theory and its
extension to the unitarity limit scattering, two questions present
themselves:
\begin{itemize}
  \item Firstly, the existence of phase coherent low-energy
  quasi-particle states in the gapless phase renders the spectral and
  (thermal) transport properties of the superconductor susceptible to
  the influence of long-range quantum interference effects. 
  \item Secondly, according to the mean-field description, a hard
  energy gap is maintained up to a critical concentration of magnetic
  impurities\footnote{\ 
  In the Born scattering limit at $T=0$, the critical concentration of
  magnetic impurities at which the energy gap goes to zero is
  $2e^{-\pi/4}\simeq 0.91$ times the critical concentration at which
  superconductivity is destroyed. In the unitarity limit, the
  critical value is decreased to
  $2 [(1 - \alpha)/(1+\alpha)]^2 e^{-\frac{\pi}{4}
  \frac{(1-\alpha)^2}{1+\alpha}}$.}. Yet, being 
  unprotected by the Anderson theorem, it would 
  seem that the gap structure predicted by the mean-field theory is
  untenable and may be destroyed by `optimal fluctuations' of the
  random impurity potential. 
\end{itemize}

In recent years, both issues have come under scrutiny. In particular,
it has been shown that the long-range, low-energy spectral and transport 
properties of a weakly disordered superconductor can be classified 
according to the their fundamental symmetries~\cite{altland_zirnbauer}.
In the quasi-classical limit $\epsilon_F\tau\gg 1$, where $1/\tau$ 
represents the scattering rate of the non-magnetic impurity distribution, 
the spectral and transport properties of the weakly disordered system can 
be presented in the framework of a statistical field theory of non-linear 
$\sigma$-model type (for a review see, e.g.~\cite{simons_altland}). Within 
this approach, the conventional Abrikosov-Gor'kov mean-field theory is 
identified as the set of (homogeneous) saddle-point equations. Mesoscopic 
fluctuations due to quantum interference effects in the particle/hole 
channel are recorded in the soft field fluctuations around the homogeneous 
mean-field solution. 

In the present case, such investigations reveal that mechanisms of 
quantum interference lead to the delocalisation of the quasi-particle
states even in low
dimension~\cite{bundschuh,senthil_fisher,read_green,bocquet_serban}. This  
behaviour provides a striking contrast with that of other superconducting 
(and normal metallic) systems where mechanism of quantum interference 
have a tendency to bring about localisation of the quasi-particle states. 
At the same time, the same general theoretical framework provides a means 
to explore the integrity of the gapped phase in the superconducting system. 
`optimal fluctuations' of the random impurity potential(s) nucleate domains
or droplets of localised tail states below the predicted mean-field gap 
edge~\cite{lamacraft_simons}. Such tail states are accommodated by 
instanton field configurations of the non-linear $\sigma$-model.

The tail states predicted by the quantum field theory differ substantially 
in character from the bound states induced by an isolated magnetic impurity. 
The former derive from mesoscopic fluctuations of the electron and hole 
wavefunctions of the normal system: specifically, in regions where the phase 
sensitivity is anomalously high, the pair-breaking effect of magnetic 
impurities (or an external magnetic field) is enhanced over that predicted
by the mean-field. Here, quasi-particle states localise over length scales
comparable to the superconducting coherence length, $\xi=(D/|\Delta|)^{1/2}$,
where $D=v_F^2\tau/d$ represents the classical diffusion constant.

As well as presenting a concise review of the field theory of the 
disordered superconducting system with magnetic impurities, the aim of 
the present paper is to explore the integrity of the sub-gap state picture
when in the presence of unitarity limit scattering. More precisely, at the 
level of mean-field, we have seen that, over a wide region of the phase 
diagram, the latter induces a delocalised band of bulk states centered 
on the bound state energy $\epsilon_{\smc{b}}$. In this case, do optimal 
fluctuations of the random potential lead to the nucleation of localised
states in the vicinity of the narrow band? 

The paper is organised as follows. In section~\ref{sec:susyf} we
refine the field theory of the weakly disordered superconducting
system to incorporate the presence of magnetic impurities. Here we
will take the magnetic impurities to be drawn from a random Poisson
distribution allowing a continuous interpolation from the unitarity
scattering limit to the Born scattering limit. Having obtained the
low-energy effective field theory, in section~\ref{sec:mfthe} we will
explore the homogeneous mean-field solution obtained from the
saddle-point of the effective action. In this case, we correctly
recover the phenomenology of Ref.~\cite{shiba} and identify the limit
in which the Born scattering Abrikosov-Gor'kov
theory~\cite{abrikosov_gorkov} is obtained. By exploring instanton
field configurations of the action, in section~\ref{sec:inhom} we
explore the integrity of the mean-field density of states. In
particular, we will show that the gap edges predicted by the
mean-field theory become mobility edges separating regions of bulk
delocalised states from localised `droplet' states generated by
optimal configurations of the random impurity potential. A brief
discussion of these results is contained within the concluding
section.

\section{Field Theory of the Superconducting System}
\label{sec:susyf}
Previous investigations have shown that, in the quasi-classical limit,
the properties of the weakly disordered superconducting system can be 
expressed in the framework of a statistical field theory of non-linear 
$\sigma$-model type. In the present case, one must consider simply how 
to tailor this analysis to the consideration of unitarity scattering 
limit of the magnetic impurity system. Since the general theoretical 
framework has been reviewed in a number of 
publications~\cite{altland_simons,bundschuh} and discussed for the 
magnetic impurity system in particular~\cite{lamacraft_simons}, we will 
keep our discussion concise focussing primarily on the idiosyncrasies of 
the present theory.

\subsection{The Model}
\label{sec:model}
In the mean-field BCS approximation, a bulk $s$-wave disordered 
superconductor in the presence of magnetic impurities is specified by
the Gor'kov Hamiltonian 
\begin{equation*}
  \hat{H}_{\smc{g}} = 
  \begin{pmatrix}
  \hat{H} & |\Delta| \sigma_2^{\smc{sp}} \\
  |\Delta| \sigma_2^{\smc{sp}} & - \hat{H}^{\mathsf{T}}
  \end{pmatrix}_{\smc{ph}} \; ,
\end{equation*}
where the index $\smc{ph}$ refers to the particle/hole space, and
Pauli matrices $\sigma^{\smc{sp}}$ operate in the spin space. Here
\begin{equation*}
  \hat{H} = \hat{\zeta}_{\hat{\vect{p}}} + V (\vect{r}) + J \vect{S}
  (\vect{r}) \cdot \vectgr\sigma^{\smc{sp}} \; .
\end{equation*}
denotes the single particle Hamiltonian, with $\hat{\zeta}_{\hat{\vect{p}}} 
= \hat{\vect{p}}^2 /2 m -\epsilon_F$, $\epsilon_F$ is the Fermi energy, 
and $|\Delta|$ is the spatially homogeneous order parameter determined from 
the self-consistency condition, $\Delta = - L^d g_\Delta \langle
\psi_{\downarrow} \psi_{\uparrow} \rangle$. In addition to a non-magnetic 
impurity potential, $V(\vect{r})$ drawn at random from a Gaussian 
white-noise impurity distribution with zero mean and variance, 
\begin{displaymath}
  \langle V (\vect{r}_1) V (\vect{r}_2) \rangle_V = \Frac{1}{2 \pi \nu 
  \tau} \delta (\vect{r}_1 - \vect{r}_2) \; ,
\end{displaymath}
the system is subjected to a \emph{classical} quenched Poisson distributed
magnetic impurity potential:
\begin{equation*}
  \vect{S} (\vect{r}) = L^d \sum_i \delta^d (\vect{r}-\vect{r}_i)
  \vect{S}_i \; ,
\end{equation*}
i.e. where the points $\vect{r}_i$ are drawn from a random Poissonian
distribution. Here, for simplicity, we will suppose that the spins
corresponding to different magnetic impurities are statistically
independent, and that the distribution over the orientation is
uniform, while the magnitude, $S$ is fixed:
i.e. $P(\{\vect{S}_i\})=\prod_i\delta (\vect{S}_i^2 - S^2)$.

Before proceeding, we should comment on the limitations of the present 
scheme. In the unitarity limit, one would expect spin $1/2$ quantum 
magnetic impurities to be fully Kondo screened by the itinerant electron
system. Our classical model is therefore limited to situations in which 
either the magnitude of the spin is sufficiently large that the moment 
can not be fully compensated or, more realistically, to systems where 
the mutual RKKY interaction of the magnetic impurities lead to a spin
glass ordering of the moments. 

\subsection{Generating Functional}
\label{sec:migfu}
To formulate a field theory of the non-interacting superconducting system,
we will follow the standard scheme~\cite{altland_simons,bundschuh} and
begin with the generating functional for the single quasi-particle Green 
function:
\begin{equation*}
  \mathcal{Z} [j] = \int D (\psi^\dag , \psi) \exp \left\{i \int d
  \vect{r} \left[\psi^\dag \left(\epsilon_+ -
  \hat{H}_{\smc{g}}\right) \psi + \psi^\dag j + j^\dag
  \psi\right]\right\} \; .
\end{equation*}
Here $\epsilon_+ = \epsilon + i0$, while $\psi^\dag (\vect{r})$ and 
$\psi (\vect{r})$ represent two independent eight component supervector 
fields with $\smc{ph}$, $\smc{sp}$ and boson/fermion ($\smc{bf}$) internal
indices. By incorporating an equal number of fermionic and bosonic 
components, the normalisation ${\cal Z}[0]=1$ is automatically imposed.
In the following, it is convenient to implement a gauge transformation
$\psi \mapsto U \psi$, where $U = E_{11}^{\smc{ph}}-E_{22}^{\smc{ph}} 
\otimes i \sigma_2^{\smc{sp}}$ and $E_{11}^{\smc{ph}} = \diag (1 , 
0)_{\smc{ph}}$, $E_{22}^{\smc{ph}} =\diag (0 , 1)_{\smc{ph}}$, whereupon, 
the Gor'kov Hamiltonian takes the simpler form:
\begin{equation*}
  \hat{H}_{\smc{g}} = \left[\hat{\zeta}_{\hat{\vect{p}}} + V
  (\vect{r})\right] \sigma_3^{\smc{ph}} + |\Delta| \sigma_2^{\smc{ph}}
  + J \vect{S} (\vect{r}) \cdot \vectgr\sigma^{\smc{sp}} \; .
\end{equation*}

As mentioned above, to determine the influence of quantum interference 
effects on the disordered superconducting system, it is useful to first 
classify the microscopic Hamiltonian according to its fundamental 
symmetries. In the absence of magnetic impurities the Gor'kov Hamiltonian 
exhibits both time-reversal symmetry, and the particle/hole symmetry
\begin{equation}
  \hat{H}_{\smc{g}} = - \sigma_2^{\smc{ph}} \otimes
  \sigma_2^{\smc{sp}} \hat{H}_{\smc{g}}^{\mathsf{T}}
  \sigma_2^{\smc{sp}} \otimes \sigma_2^{\smc{ph}} \; .
\label{eq:symcc}
\end{equation}
In presence of magnetic impurities the time-reversal is broken, while
the particle/hole symmetry is conserved. Applied to the corresponding
Gor'kov Green function, $\hat{G}_{\smc{g}}^{\smc{r,a}} (\epsilon) =
(\epsilon_{\pm} -\hat{H}_{\smc{g}})^{-1}$, the particle/hole
transformation~\eqref{eq:symcc} converts an advanced function into a
retarded one:
\begin{equation*}
  \hat{G}_{\smc{g}}^{\smc{r,a}} (\epsilon) = - \sigma_2^{\smc{ph}}
  \otimes \sigma_2^{\smc{sp}} \left[\hat{G}_{\smc{g}}^{\smc{a,r}}
  (-\epsilon)\right]^{\mathsf{T}} \sigma_2^{\smc{sp}} \otimes
  \sigma_2^{\smc{ph}} \; .
\end{equation*}

As usual~\cite{efetov,bundschuh}, to accommodate quantum interference
effects in the particle/hole channel, it is convenient to affect a
further space doubling $\psi^\dag\hat{G}_{\smc{g}}^{\smc{r}}
(\epsilon) \psi = \bar{\Psi}\hat{G}^{\smc{r}}_{\smc{g}} (\epsilon
\sigma_3^{\smc{cc}}) \Psi$, where, in the charge conjugation space,
the vector fields take the form\footnote{\ 
  The transposition operation for the supervectors $\psi$ and
  $\psi^\dag$ and the supermatrix $F$ is chosen 
  according to the convention: 
  \begin{align*}
  \psi^\mathsf{T} &= \begin{pmatrix}
  \phi & \chi
  \end{pmatrix}_{\smc{bf}} & {\psi^\dag}^\mathsf{T} &=
  \begin{pmatrix}
  \phi^* \\ - \chi^*
  \end{pmatrix}_{\smc{bf}} &
  F^{\mathsf{T}} &= \begin{pmatrix}
  a & \rho \\
  -\sigma & b
  \end{pmatrix}_{\smc{bf}} \; .
  \end{align*}}:
\begin{align*}
  \Psi &= \Frac{1}{\sqrt{2}} 
  \begin{pmatrix} \psi \\
  \sigma_2^{\smc{ph}} \otimes \sigma_2^{\smc{sp}}
  {\psi^\dag}^{\mathsf{T}}
  \end{pmatrix}_{\smc{cc}} &
  \bar{\Psi} &= \frac{1}{\sqrt{2}} 
  \begin{pmatrix} \psi^\dag & -\psi^{\mathsf{T}} \sigma_2^{\smc{sp}}
  \otimes \sigma_2^{\smc{ph}}  
  \end{pmatrix}_{\smc{cc}} \; .
\end{align*}
This completes the formulation of the generating functional for the
single particle properties of the Gor'kov Hamiltonian. The theory is 
specified in terms of $16$-component supervector fields $\Psi$ and
$\bar{\Psi}$ with the following symmetry relations.
\begin{align*}
  \Psi &= - \sigma_2^{\smc{ph}} \otimes \sigma_2^{\smc{sp}} \gamma 
  \bar{\Psi}^{\mathsf{T}} & \bar{\Psi} &= \Psi^{\mathsf{T}}
  \sigma_2^{\smc{ph}} \otimes \sigma_2^{\smc{sp}} \gamma^{\mathsf{T}}
 \; ,
\end{align*}
with $\gamma = i\sigma_2^{\smc{cc}}  E^{\smc{bf}}_{11} - \sigma_1^{\smc{cc}} 
E^{\smc{bf}}_{22}$. As before $E^{\smc{bf}}_{11} = \diag(1,0)_{\smc{bf}}$ and 
$E^{\smc{bf}}_{22} = \diag (0,1)_{\smc{bf}}$ represent projection operators
on the boson/fermion space.

\subsection{(Non-Magnetic) Impurity Averaging}
\label{sec:nonma}
An ensemble average of the generating functional over the non-magnetic 
impurity distribution $V$ induces a quartic interaction of the fields 
which can be decoupled by means of a Hubbard-Stratonovich transformation 
with the introduction of $16\times 16$ supermatrix fields $Q(\vect{r})$ 
\begin{equation*}
  \langle \int \exp \left[ - i \int d \vect{r} \; \bar{\Psi} V
  \sigma_3^{\smc{ph}} \Psi \right] \rangle_V 
  = \int D Q \exp \left[ \int d \vect{r} \left( \Frac{\pi \nu}{8 
  \tau} \str \; Q^2 - \Frac{1}{2 \tau} \bar{\Psi} Q 
  \sigma_3^{\smc{ph}} \Psi \right) \right] \; .
\end{equation*}
The symmetry properties of $Q$ are inherited from the dyadic
product $\sigma_3^{\smc{ph}} \Psi (\vect{r}) \otimes \bar{\Psi}
(\vect{r}$) and impose the condition
\begin{equation}
  Q = \sigma_1^{\smc{ph}} \otimes \sigma_2^{\smc{sp}} \gamma \;
  Q^{\mathsf{T}} \gamma^{\mathsf{T}} \sigma_2^{\smc{sp}} \otimes
  \sigma_1^{\smc{ph}} \; .
\label{eq:symme}
\end{equation}

In principle we could immediately subject the generating functional to 
a further average over the Poisson distributed magnetic impurity potential. 
However, such an approach proves to be unprofitable. Since the typical
separation of magnetic impurities, $\ell_s=v_F\tau_s$, is greatly in
excess of the mean-free path associated with the non-magnetic
impurities, $\ell = v_F \tau$, it is more sensible to postpone the 
second ensemble average until the quasi-classical theory has been 
developed. Therefore, at this stage, let us proceed by integrating out 
the superfields after which the generating functional assumes the form
\begin{equation}
  \langle {\mathcal Z} [0] \rangle_{V} = \int DQ \exp 
  \left\{ \Frac{\pi \nu}{8 \tau} \int d \vect{r} \; \str Q^2
  (\vect{r}) - \Frac{1}{2} \int d \vect{r} \; \str \langle \vect{r} |
  \ln \hat{{\mathcal{G}}}^{-1} | \vect{r} \rangle \right\} \; ,
\label{eq:zgfun}
\end{equation}
where $\hat{{\mathcal{G}}}$ represents the supermatrix Green
function,
\begin{align*}
  \hat{{\mathcal{G}}}^{-1}   &= \hat{{\mathcal{G}}}_0^{-1} + \epsilon
  \sigma_3^{\smc{cc}} - |\Delta| \sigma_2^{\smc{ph}} - J \vect{S}
  \cdot \vectgr\sigma^{\smc{sp}}, &
  \hat{{\mathcal{G}}}_0^{-1} &= i0\sigma_3^{\smc{cc}} -
  \hat{\zeta}_{\hat{\vect{p}}} \sigma_3^{\smc{ph}} + \Frac{i}{2 \tau} Q
  \sigma_3^{\smc{ph}} \; .
\end{align*}

\subsection{Intermediate Energy Saddle-Point: the Non-Linear
$\sigma$-Model} 
\label{sec:firsp}
To make further progress it is necessary to employ a saddle-point 
approximation. Following Ref.~\cite{altland_simons}, it is convenient to 
implement a two-step procedure making use of the hierarchy of energy 
scales which place the superconductor in the quasi-classical and dirty
limits: 
\begin{displaymath}
  \epsilon_F \gg \frac{1}{\tau} \gg \{\frac{1}{\tau_s} , 
  |\Delta|\} \gg \delta\; .
\end{displaymath}
To implement the quasi-classical approximation, we therefore temporarily 
suspend the energy source, $\epsilon$, the order parameter, $|\Delta|$ and
the magnetic impurity potential, $J \vect{S}$, and seek an intermediate 
energy scale saddle-point.
Such an analysis is discussed in detail in the literature~\cite{efetov}
and here we only recapitulate the results. A variation of the action with 
respect to $Q$ obtains the saddle-point equation
\begin{displaymath}
  Q(\vect{r}) = \Frac{i}{\pi \nu} \sigma_3^{\smc{ph}} 
  \langle \vect{r} |\hat{{\mathcal{G}}}_0 | \vect{r} \rangle\; .
\end{displaymath}
Taking into account the analytical properties of the Green function, in
the pole approximation, one obtains the conventional saddle-point solution 
$Q_{\text{sp}} = \sigma_3^{\smc{ph}}\otimes\sigma_3^{\smc{cc}}$. However, 
the saddle-point solution is not unique but spans the non-linear manifold 
$Q^2 = \mathbb{I}$: in the absence of the external symmetry breaking 
perturbations, the saddle-point equation admits an entire manifold of 
homogeneous solutions parameterised by transformations $Q=TQ_{\text{sp}} 
T^{-1}$, where $T$ is a supermatrix, constant in space, and compatible with 
the charge conjugation symmetry properties of $Q$, \eqref{eq:symme}. 

Fluctuations of $Q$ transverse to the saddle-point manifold are massive 
and may be integrated out within the saddle-point approximation, justified 
by the large parameter $1/\tau \delta$. By contrast, fluctuations which
preserve the non-linear constraint $Q^2 = \mathbb{I}$ are massless and 
must be integrated exactly. In the same quasi-classical approximation 
$\epsilon_F \tau \gg 1$, the matrix Green function takes the form,
\begin{equation*}
  \mathcal{G}_0 (\vect{r}_1 , \vect{r}_2) = - i \pi \nu f_d
  (|\vect{r}_1 - \vect{r}_2|) \sigma_3^{\smc{ph}} Q
  \left(\Frac{\vect{r}_1 + \vect{r}_2}{2}\right) \; ,
\end{equation*}
where the Friedel function $f_d(r)=\langle \Im G^{\smc{a}} (\vect{r} ,
0) \rangle_V / \langle \Im G^{\smc{a}}_0 (0,0) \rangle_V$ denotes the
impurity averaged single-particle Green function.

To determine the intermediate energy scale action, we now restore the 
symmetry breaking parameters $\epsilon$, $|\Delta|$ and $J \vect{S}$.
Expanding the term $\str \ln\hat{{\mathcal{G}}}^{-1}$ appearing in 
\eqref{eq:zgfun}, to leading order in $\epsilon$ and $|\Delta|$, one 
obtains:
\begin{equation}
  \ln \hat{{\mathcal{G}}}^{-1} \simeq \ln \hat{{\mathcal{G}}}_0^{-1} +
  \left[ \hat{{\mathcal{G}}}_0 \left( \epsilon \sigma_3^{\smc{cc}} -
  |\Delta| \sigma_2^{\smc{ph}} \right) \right] + \ln \left( \mathbb{I} -
  \hat{{\mathcal{G}}}_0 J \vect{S} \cdot \vectgr\sigma^{\smc{sp}}
  \right) \; ,
\label{eq:expan}
\end{equation}
where, importantly, the magnitude of the magnetic impurity potential has 
been left unrestricted and where the neglected terms turn out to be of
order $\epsilon \tau \ll 1$ or $|\Delta| \tau \ll 1$. Taking into
account slow fluctuations $Q(\vect{r}) = T^{-1}
(\vect{r})\sigma_3^{\smc{ph}} \otimes \sigma_3^{\smc{cc}} T
(\vect{r})$, a gradient expansion of the first two terms of the
series~\eqref{eq:expan} to leading order in $|\Delta| \tau$ and
$\epsilon \tau$ recovers the familiar non-linear $\sigma$-model action
for the disordered superconductor,
\begin{equation}
  S_0 [Q] = -\frac{\pi \nu}{8} \int d \vect{r} \; \str \left[D (\nabla
  Q)^2 + 4  \left(i \epsilon_+ \sigma_3^{\smc{ph}} \otimes
  \sigma_3^{\smc{cc}} + |\Delta| \sigma_1^{\smc{ph}}\right) Q\right]
  \; ,
\label{eq:barea}
\end{equation}
where $D = v_F^2 \tau / d$ represents the classical diffusion constant 
associated with the non-magnetic impurities. Taken together with the spin 
scattering contribution, the average generating functional assumes the form
\begin{gather*}
  \langle \mathcal{Z} [0] \rangle_V = \int_{Q^2 = \mathbb{I}} DQ 
  e^{- S_0[Q] - S_S[Q]}
\intertext{where}
  S_S [Q] = \Frac{1}{2} \int d \vect{r} \; \str \langle \vect{r} |\ln
  \left( \mathbb{I} - \hat{{\mathcal G}}_0 J \vect{S} \cdot
  \vectgr\sigma^{\smc{sp}}\right) | \vect{r} \rangle \; .
\end{gather*}
To make sense of the spin scattering component of the action, it is
now necessary to implement the magnetic impurity average.

\subsection{Magnetic Impurity Averaging}
\label{sec:magne}
If we restricted ourselves to the limit of Born scattering, we would be 
free to expand the action $S_S[Q]$ to second order\footnote{\
  Note that the term linear in $\vect{S}$ generated by the expansion
  vanishes when projected onto the singlet
  saddle-point~\eqref{eq:mians}, considered below.} 
in $\vect{S}$. In this case, once ensemble averaged over the magnetic 
impurity distribution, we would obtain the weak coupling action $S[Q] 
\simeq S_0[Q] + S_S^{\smc{ba}}[Q]$, where
\begin{equation*}
  S_S^{\smc{ba}} = \Frac{n_s (\pi \nu L^d J S)^2}{4d_n} \int d
  \vect{r} \; \str \left[ \sigma_3^{\smc{ph}} \otimes
  \vectgr\sigma^{\smc{sp}} Q (\vect{r}) \right]^2 \; .
\end{equation*}
With the identification ($d_n = 3$)
\begin{equation}
  2 \alpha n_s = \Frac{\pi \nu}{\tau_s} \; ,
\label{eq:ident}
\end{equation}
where $1/\tau_s$ represents the magnetic impurity scattering rate, 
$\alpha = (\pi \nu J L^d S)^2$, and $n_s$ is the magnetic impurity
concentration, this result coincides with that obtained by 
Ref.~\cite{lamacraft_simons}. However, in the unitarity scattering limit, 
we are not at liberty to freely take the impurity potential from underneath 
the logarithm. 

To simplify the action $S_S[Q]$, let us recall that the typical separation
of magnetic impurities, $\ell_s$, is greatly in excess of
the mean-free path associated with the non-magnetic impurities,
$\ell$. (In the opposite limit, superconductivity would, in any 
case, be fully suppressed.) In this case, we may affect the approximation
\begin{equation*}
  {\mathcal{G}}_0 (\vect{r}_i , \vect{r}_j) \simeq - i \pi \nu 
  \sigma_3^{\smc{ph}} Q(\vect{r}_i) \delta_{ij} \; .
\end{equation*}
Making use of this relation, an expansion and re-summation of the logarithm 
leads to the result:
\begin{displaymath}
  S_S [Q] \simeq \frac{1}{2} \sum_i \str \ln \left[ \mathbb{I} + (i \pi
  \nu J L^d) \sigma_3^{\smc{ph}} Q (\vect{r}_i) \vect{S}_i \cdot
  \vectgr\sigma^{\smc{sp}} \right] \; .
\end{displaymath}

In this approximation, the functional has become separable in the individual 
magnetic impurity scatterers. In principle, we could proceed directly by 
subjecting the generating functional to an ensemble average over the random 
Poissonian distributed magnetic impurity distribution. However, for a general 
supermatrix $Q$, without expanding the logarithm, the average over the 
independent spin degrees of freedom is laborious and not illuminating. 
Instead, we will follow a different program.

Firstly, by taking into account the symmetry of the soft degrees of freedom, 
we can further simplify the analysis: specifically, from the previous 
analysis of the Born scattering theory, it is evident that fluctuations of 
$Q$ which are not proportional to $\mathbb{I}^{\smc{sp}}$ will be rendered 
massive by the magnetic impurity potential. Since we are interested in the 
low-energy content of the theory, we may therefore specialise our 
considerations to the singlet degrees of freedom of $Q$, i.e. $Q\mapsto
Q\otimes\mathbb{I}^{\smc{sp}}$. In this case, if we undertake the trace 
over the spin degrees of freedom of $Q$, one obtains
\begin{equation*}
  S_S [Q] = \Frac{1}{2} \sum_i \str_8 \ln \left[\mathbb{I} - \alpha_i
  \sigma_3^{\smc{ph}} Q(\vect{r}_i) \sigma_3^{\smc{ph}}
  Q(\vect{r}_i)\right] \; , 
\end{equation*}
where $\alpha_i = (\pi \nu J L^d |\vect{S}_i|)^2 \mapsto \alpha$,
constant, is the dimensionless scattering amplitude associated with
each magnetic impurity, and the notation $\str_8$ indicates that we
have carried out the trace over the spin indices.

Secondly, the structure of the non-linear $\sigma$-model 
action~\eqref{eq:barea} shows that the dominant contributions to the
generating functional arise from field configurations which vary
slowly at the scale of the coherence length. Therefore, in the dense
magnetic impurity limit, where the typical separation between magnetic
impurities, $\ell_s=v_F\tau_s$, is much smaller than $\xi$, one can
expect the density of spin scatterers, $\rho (\vect{r}) = \sum_i
\delta (\vect{r} - \vect{r}_i)$, to be dominated by it's smooth
average\footnote{\
  Such an approximation circumvents the need to implement the
  averaging over the Poisson distribution explicitly. Later, we will
  see that this brings with it considerable simplification in the
  instanton analysis, while leaving the mean-field analysis
  unchanged.},
$n_s$. In this approximation, one obtains
\begin{equation*}
  \widetilde{S}_S [Q] = \Frac{n_s}{2} \int d\vect{r}
  \str_8 \ln \left[\mathbb{I} - \alpha\,
  \sigma_3^{\smc{ph}} Q(\vect{r}) \sigma_3^{\smc{ph}}
  Q(\vect{r})\right] \; .
\end{equation*}
Notice that the invariance properties of $Q$ on the saddle-point manifold 
imply a symmetry of the action under the transformation $\alpha \mapsto 
1/\alpha$: the weak (or Born) scattering limit is therefore `dual' to the 
strong scattering limit. 

This concludes the construction of the field theory of the disordered 
superconducting system. Single quasi-particle properties of the
superconductor are presented as a functional field integral $\langle \dots 
\rangle_Q=\int DQ \dots e^{-S_{\text{eff}}[Q]}$ involving the effective 
non-linear $\sigma$-model action 
\begin{equation*}
  S_{\text{eff}}[Q] = S_0 [Q] + \widetilde{S}_S [Q] \; .
\end{equation*}
In particular, the local quasi-particle DoS can be obtained from the 
functional integral 
\begin{equation*}
  \langle \nu (\epsilon , \vect{r}) \rangle_{V,S} = \Frac{\nu}{16} \Re 
  \langle \str_8 \left[\sigma_3^{\smc{bf}} \otimes
  \sigma_3^{\smc{ph}} \otimes \sigma_3^{\smc{cc}} Q (\vect{r})\right]
  \rangle_Q \; .
\end{equation*}

Although the soft mode action is stabilised by the large parameter
$\epsilon_F\tau\gg 1$, the majority of field fluctuations of the
action are rendered massive: the energy source $\epsilon$, the order
parameter $|\Delta|$ and the magnetic impurity potential all lower the
symmetry of the intermediate energy scale theory. To assimilate the effect 
of these terms, and to establish contact with the Abrikosov-Gor'kov 
theory~\cite{abrikosov_gorkov} in the Born approximation limit,
and with Ref.~\cite{shiba} in the unitarity limit, in the following 
section, we will proceed by exploring the low-energy saddle-point 
structure of the theory.

\subsection{Low-Energy Saddle-Point}
\label{sec:mfthe}
To explore the low-energy structure of the theory we proceed by varying the
action $S_{\text{eff}} [Q]$ with respect to $Q$ subject to the non-linear
constraint $Q^2=\mathbb{I}$. In doing so, one obtains the saddle-point 
equation
\begin{equation*}
  D \nabla (Q \nabla Q) + [i \epsilon_+ \sigma_3^{\smc{ph}} \otimes 
  \sigma_3^{\smc{cc}} + |\Delta| \sigma_1^{\smc{ph}} , Q] -
  \Frac{1}{2\tau_s} [ \Frac{1}{\mathbb{I} - \alpha \sigma_3^{\smc{ph}}
  Q \sigma_3^{\smc{ph}} Q} \sigma_3^{\smc{ph}} , Q \sigma_3^{\smc{ph}}
  Q ] = 0 \; ,
\end{equation*}
where the scattering rate $1/\tau_s$ is related to the magnetic impurity 
concentration $n_s$ and the scattering amplitude $\alpha$ through the 
relation~\eqref{eq:ident}. 
The latter must be supplemented by the self-consistency condition
\begin{displaymath}
  \Frac{1}{g_\Delta}|\Delta| = \Frac{\pi}{8\beta \delta} \left. \sum_n
  \str_8 \left[\sigma_1^{\smc{ph}} \otimes \sigma_3^{\smc{bf}}
  Q\right]\right|_{i\epsilon_+ = -\epsilon_n} \; ,
\end{displaymath}
where $g_\Delta$ is the BCS coupling constant and $Q$ represents the 
solution of the mean-field equation taken at the Matsubara frequencies 
$\epsilon \mapsto i \epsilon_n = i (2n + 1)\pi/\beta$. Applying the Ansatz 
for the saddle-point,
\begin{equation}
  Q (\vect{r}) = \sin \hat{\theta} (\vect{r})
  \sigma_1^{\smc{ph}} + \cos \hat{\theta} (\vect{r})
  \sigma_3^{\smc{ph}} \otimes \sigma_3^{\smc{cc}} \; ,
\label{eq:mians}
\end{equation}
where the matrix $\hat{\theta} = \diag ( i \theta_{\smc{bb}} ,
\theta_{\smc{ff}})_{\smc{bf}}$ is diagonal in the superspace and
independent of the other indices, one obtains the saddle-point 
equation
\begin{equation}
  D \nabla^2 \hat{\theta} + 2  \left( i \epsilon \sin \hat{\theta} -
  |\Delta| \cos \hat{\theta} \right) - \Frac{1}{\tau_s} \Frac{\sin 2
  \hat{\theta}}{\mathbb{I} + \alpha^2 + 2 \alpha \cos 2 \hat{\theta}}
  = 0 \; .
\label{eq:3eqnm}
\end{equation}

As expected, in the limit $\alpha \ll 1$, an expansion of the spin 
scattering term obtains the Born scattering equation analysed in 
Ref.~\cite{lamacraft_simons}. Here, taking the saddle-point solution to 
be homogeneous in space and symmetric in the superspace, $i\theta_{\smc{bb}}
=\theta_{\smc{ff}}\equiv \theta$, one recovers the Abrikosov-Gor'kov
mean-field equations. Similarly, in the strong scattering limit 
$\alpha\gg 1$, the action coincides with that obtained in the Born 
scattering limit. This coincidence reflects the duality seen on the 
level of the action and can be understood qualitatively in the following 
way: when the local spin scattering potential is very strong, the 
wavefunction for the low-energy quasi-particles states is strongly 
suppressed at the impurity centre. As a result, the matrix element for 
the effect spin scattering rate is itself strongly suppressed. 
However, our main interest is in the crossover region where the dimensionless 
scattering rate is $\alpha \sim 1$. Here the saddle-point 
equation is given by~\eqref{eq:3eqnm}.

\begin{figure}
\begin{center}
\includegraphics[width=1\linewidth,angle=0]{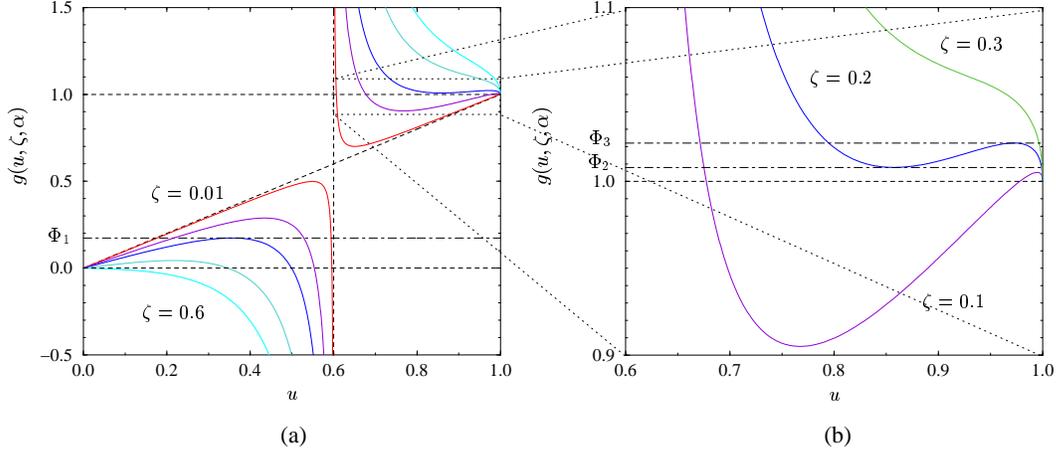}
\end{center}
\caption{\small
        Function $g(u,\zeta,\alpha)$ versus $u$; (a) for $\alpha=0.25$ 
        and for $\zeta = 0.01$, $\zeta = 0.1$, $\zeta = 0.2$,
        $\zeta = 0.4$ and $\zeta = 0.6$. The value of $\Phi_1 =
        \epsilon_g / |\Delta|$ is explicitly indicated for the case
        $\zeta = 0.2$;
        (b) the function $g(u,\zeta,\alpha)$ is plotted in the region
        $u > \gamma$ for $\alpha=0.25$ and for $\zeta = 0.1$, $\zeta =
        0.2$ and $\zeta = 0.3$. The values of the edges $\Phi_2$ and
        $\Phi_3$ are explicitly indicated for the case $\zeta = 0.2$.
        }
\label{fig:gfunc}
\end{figure}

As with the Born scattering limit, to analyse the saddle-point equations,
we first seek a homogeneous supersymmetric solution, $\hat{\theta}_{\smc{mf}} 
= \theta_{\smc{mf}} \mathbb{I}^{\smc{bf}}$. Then, defining a `renormalised' 
energy and order parameter,
\begin{equation}
\begin{split}
  \tilde{\epsilon} &= \epsilon + \Frac{i}{2 \tau_s} \Frac{\cos 
  \theta_{\smc{mf}}}{1 + \alpha^2 + 2 \alpha \cos 2 \theta_{\smc{mf}}}
  \\ 
  |\tilde{\Delta}| &= |\Delta| + \Frac{1}{2 \tau_s} \Frac{\sin
  \theta_{\smc{mf}}}{1 + \alpha^2 + 2 \alpha \cos 2 \theta_{\smc{mf}}}
  \; ,
\end{split}
\label{eq:ghomo}
\end{equation}
the homogeneous saddle-point equation~\eqref{eq:3eqnm} takes a conventional
BCS form, $i\tilde{\epsilon} \sin \theta_{\smc{mf}} = |\tilde{\Delta}| \cos
\theta_{\smc{mf}}$. Defining the dimensionless parameter $u =\tilde{\epsilon} 
/ |\tilde{\Delta}|$ such that
\begin{align*}
  \cos \theta_{\smc{mf}} &= \Frac{-iu}{\sqrt{1 - u^2}} &
  \sin \theta_{\smc{mf}} &= \Frac{-1}{\sqrt{1 - u^2}} \; ,
\end{align*}
the mean-field equation can be cast in the following form:
\begin{equation}
  \Frac{\epsilon}{|\Delta|} = g(u,\zeta,\alpha) \equiv u \left( 1 -
  \zeta_\alpha \Frac{\sqrt{1 - u^2}}{\gamma^2 - u^2} \right) \; ,
\label{eq:gufun}
\end{equation}
where $\zeta_\alpha = \zeta / (1 + \alpha)^{2}$ and $\gamma = |1 -
\alpha| / (1 + \alpha)$. As expected, despite the inclusion of an
additional non-magnetic impurity potential, this result coincides with
that obtained in Ref.~\cite{shiba} by diagrammatic expansion in the
T-matrix approximation. This equivalence occurs because the nature of
the electron motion (diffusive or ballistic) has no effect at the
homogeneous mean-field level. By contrast, soon we will see that the
nature of the underlying electron dynamics does impact on the nature
of the fluctuations around the saddle-point.

The homogeneous saddle-point solutions are given by the values of $u$
satisfying Eq.~\eqref{eq:gufun}. As in the Abrikosov-Gor'kov theory, a
number of features of the DoS can be deduced
starting from the shape of the function $g(u,\zeta,\alpha)$ when the
impurity concentration $\zeta$ and the scattering amplitude $\alpha$
are varied. In Figure~\ref{fig:gfunc} the typical dependence of
$g(u,\zeta,\alpha)$ in $u$ is shown. The parameter $\gamma$ defines the
second order pole, while for values of $\zeta$ sufficiently low the
function shows one extremum in the region $u \in [0 , \gamma)$ and the
two extrema in the region $u \in (\gamma , 1]$, that disappear as the
value of the impurity concentration increases. As will be clear
later, the three extrema of $g$ have a clear  interpretation in terms
of the sharp edges of the DoS.

As in the Born limit ($\alpha = 0$, $\gamma = 1$), the maximum of $g
(u , \zeta , \alpha)$ in the interval $u \in [0 , \gamma)$ has the
meaning of the rescaled energy gap, $\Phi_1 = \epsilon_g /
|\Delta|$, so that the set of the values of $\alpha$ and $\zeta$ for
which $\epsilon_g$ becomes zero defines the gapless region for the
superconductor in the unitarity limit:
\begin{align*}
  \left. g' (u , \zeta , \alpha) \right|_{u = 0} &= 0 & {}&
  \Rightarrow & \zeta &= \zeta_0 \equiv (1 - \alpha)^2 \; .
\end{align*}
This means that in the region $\zeta \ge (1 - \alpha)^2$ the superconductor 
is gapless ($\epsilon_g = 0$, $|\Delta| \ne 0$). Compared with the Born 
limit, this result shows that the concentration at which the system enters 
the gapless phase is renormalised by the finite scattering amplitude $\alpha$.

\begin{figure}
\begin{center}
\includegraphics[width=1\linewidth,angle=0]{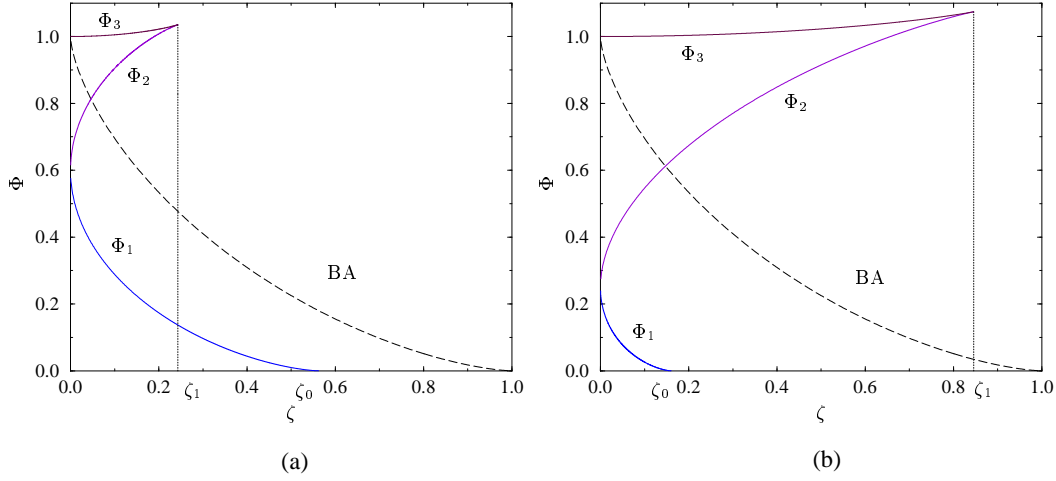}
\end{center}
\caption{\small Rescaled energy gap $\epsilon_g / |\Delta|=
        \Phi_1$, edge of the impurity band $\Phi_2$ and edge
        of the continuum  $\Phi_3$ versus $\zeta$; (a) for $\alpha =
        0.25$, $\zeta_1 < \zeta_0 = (1-\alpha)^2 \simeq 0.56$;  
        (b) for $\alpha = 0.6$, $\zeta_1 > \zeta_0 =0.16$. The dashed
        line is the energy gap in the Born approximation case, that in
        this case coincides with the edge of the continuum.}
\label{fig:phiun}
\end{figure}

The two extrema in the region $u \in (\gamma , 1]$, which we denote
$\Phi_2$ and $\Phi_3$ (see~Fig.~\ref{fig:gfunc} (b)), represent
respectively the top of the impurity band and the bottom of the
continuum of bulk states. The dependence of $\epsilon_g / |\Delta|$,
$\Phi_2$ and $\Phi_3$ on $\zeta$ is shown in Figure~\ref{fig:phiun}
for two representative values of $\alpha$. Defining $\zeta_0$ and 
$\zeta_1$ respectively as the value of $\zeta$ at which the energy gap 
becomes zero and the value of $\zeta$ at which the top of the impurity
band touches the bottom of the continuum, when $\zeta_1 < \zeta_0$
(Fig.~\ref{fig:phiun} (a)) the top of the impurity band touches the
bottom of the continuum before the superconductor becomes gapless,
while the opposite situation pertains when $\zeta_0 < \zeta_1$
(Fig.~\ref{fig:phiun} (b)).

The assigned definitions of $\epsilon_g / |\Delta|$, $\Phi_2$ and
$\Phi_3$ become clear when we plot the normalised mean-field density
of states $\langle \nu_{\smc{mf}} (\epsilon) \rangle_{V,S} / \nu$ as a
function of the rescaled energy, $\epsilon/|\Delta|$: 
\begin{equation*}
  \langle \nu_{\smc{mf}} (\epsilon) \rangle_{V,S} = \Frac{\nu}{16} \Re
  \langle \str \left[\sigma_3^{\smc{bf}} \otimes \sigma_3^{\smc{ph}}
  \otimes \sigma_3^{\smc{cc}} Q_{\smc{mf}}\right] \rangle_Q
  = \nu \Re \cos \theta_{\smc{mf}} = \Im \left(\Frac{u}{\sqrt{1 -
  u^2}}\right) \; .
\end{equation*}
In order to explicitly find $\langle \nu_{\smc{mf}} (\epsilon)
\rangle_{V,S} / \nu$, we have to solve equation~\eqref{eq:gufun},
expressing $u$ in terms of $\epsilon / |\Delta|$. For values of
$\alpha$ and $\zeta$ at which all the edges, $\epsilon_g / |\Delta| =
\Phi_1$, $\Phi_2$ and $\Phi_3$ are different from zero,
equation~\eqref{eq:gufun} has complex solutions for $\Phi_1 <
\epsilon/|\Delta| < \Phi_2$ and for $\epsilon/|\Delta| > \Phi_3$. For
this reason, in this case, the DoS of a single particle excitation
consists of two distinct parts: over the energy range $\Phi_1 <
\epsilon/|\Delta| < \Phi_2$ the system exhibits an impurity band while
for $\epsilon/|\Delta| > \Phi_3$ there exists a continuum of bulk
states. Within the intervening energy intervals, the mean-field DoS is
zero. As $\zeta$ increases, the energy gap closes, while the top of
the impurity band converges on the edge of the continuum. Which
happens first depends on the value of the magnetic scattering
amplitude $\alpha$. Figure~\ref{fig:dos1u} clearly shows the
phenomenon of the impurity band growth around the energy
\begin{displaymath}
  \Frac{\epsilon_{\smc{b}}}{|\Delta|} = \gamma = \Frac{|1 - \alpha|}{1
  + \alpha} \; ,
\end{displaymath}
corresponding to that of the bound state developed around a single 
isolated magnetic impurity~\cite{shiba,salkola_balatsky,flatte_byers}.

This concludes our discussion of the solutions of the homogeneous
mean-field equations and their ramifications on the DoS. However, as
mentioned in the introduction, the integrity of a quasi-particle
energy gap (or gaps) predicted by the mean-field theory does not 
seem tenable. Following the discussion of the Born scattering 
system~\cite{lamacraft_simons}, we expect optimal fluctuations of the 
random impurity potential to generate fluctuations in the effective spin 
scattering rate which in turn must lead to the nucleation of localised 
tail states which soften the gap edge. In Ref.~\cite{lamacraft_simons} 
it was shown that such localised sub-gap quasi-particle states are reflected 
in spatially inhomogeneous instanton field configurations of the 
$\sigma$-model action. In the present case, we expect an analogous situation 
to persist even in the unitarity scattering limit: however, in this case, 
we must expect that all three of the hard edges $\Phi_1$, $\Phi_2$ and 
$\Phi_3$ are softened by the bounce configurations. 

\section{Instantons and Tail States}
\label{sec:inhom}
To explore the structure of the tail state distribution, it is
necessary to revisit the saddle-point equation~\eqref{eq:3eqnm} and
look for inhomogeneous solutions. For this purpose, it is convenient
to recast the saddle-point equation~\eqref{eq:3eqnm} in terms of its
first integral
\begin{gather*} 
  D (\nabla \hat{\theta})^2 - |\Delta| V (\hat{\theta}) = \text{const}
  \; ,
\intertext{where}
  V (\hat{\theta}) = 4 \left(i \Frac{\epsilon}{|\Delta|} \cos
  \hat{\theta} + \sin \hat{\theta}\right)- \Frac{\zeta}{2 \alpha}
  \left[\ln \left(\mathbb{I} + \alpha^2 + 2 \alpha \cos
  \hat{\theta}\right) - \mathbb{I}\right] \; ,
\end{gather*}
represents the effective complex potential. The identification of the 
saddle-point solution is simplified by the observation that the homogeneous 
mean-field configuration~\eqref{eq:ghomo} satisfies the condition 
$\theta_{\smc{mf}} (\epsilon < \epsilon_g) =-\pi /2 -i \phi_{\smc{mf}}$, 
with $\phi_{\smc{mf}}$ real (i.e. so that the mean-field DoS, $\langle 
\nu_{\smc{mf}} (\epsilon) \rangle_{V,S} = \nu\Re \cos \theta_{\smc{mf}}$, 
vanishes below the energy gap). One can therefore identify a `bounce' 
solution parameterised by $\theta = -\pi /2 - i \phi$, with $\phi$ real, 
and involving the \emph{real} potential $V_{\smc{r}} (\phi) \equiv V ( 
- \pi /2 - i \phi)$. 

\begin{figure}
\begin{center}
\includegraphics[width=0.6\linewidth,angle=0]{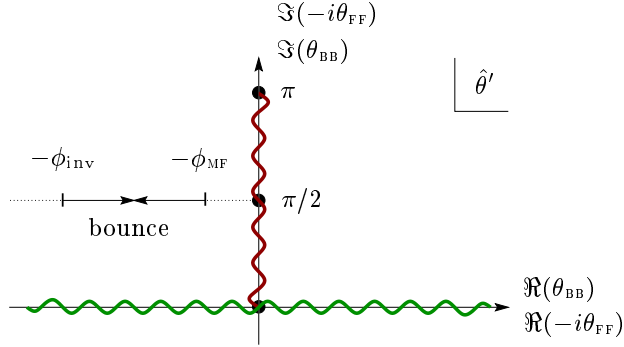}
\end{center}
\caption{\small
        Integration contours for boson-boson and fermion-fermion
        fields in the complex $\hat{\theta'} = - i \hat{\theta}$
        plane. 
        }
\label{fig:conto}
\end{figure}

Now integration over the angles $\hat{\theta}=\diag ( i
\theta_{\smc{bb}} , \theta_{\smc{ff}})_{\smc{bf}}$ is constrained to
certain contours~\cite{efetov}: the contour of integration over
the boson-boson field $\theta_{\smc{bb}}$ includes the entire real
axis, while the fermion-fermion field $i \theta_{\smc{ff}}$ runs along the
imaginary axis from $0$ to $i \pi$ (see Fig.~\ref{fig:conto}). As the
saddle-point solution must be accessible by a smooth deformation of the 
integration contour, the bounce solution is accessible only to the 
boson-boson field, 
\begin{align}
  i \theta_{\smc{bb}} (\vect{r}) &= - \Frac{\pi}{2} - i \phi (\vect{r}) &
  \theta_{\smc{ff}} &= \theta_{\smc{mf}} \; ,
\label{eq:bounc}
\end{align}
i.e. the particular bounce configuration is non-trivial in the superspace.
With this parameterisation, the saddle-point equation assumes the form
$(\nabla_{\vect{r} / \xi} \phi)^2 + V_{\smc{r}} (\phi) = V_{\smc{r}}
(\phi_{\smc{mf}})$, where 
\begin{equation}
  V_{\smc{r}} (\phi) = 4 \left(\Frac{\epsilon}{|\Delta|} \sinh \phi -
  \cosh \phi\right)- \Frac{\zeta}{2 \alpha} \left[\ln\left|1 +
  \alpha^2 - 2 \alpha \cosh 2 \phi\right| - 1\right] \; .
\label{eq:fpote}
\end{equation}

\begin{figure}
\begin{center}
\includegraphics[width=1\linewidth,angle=0]{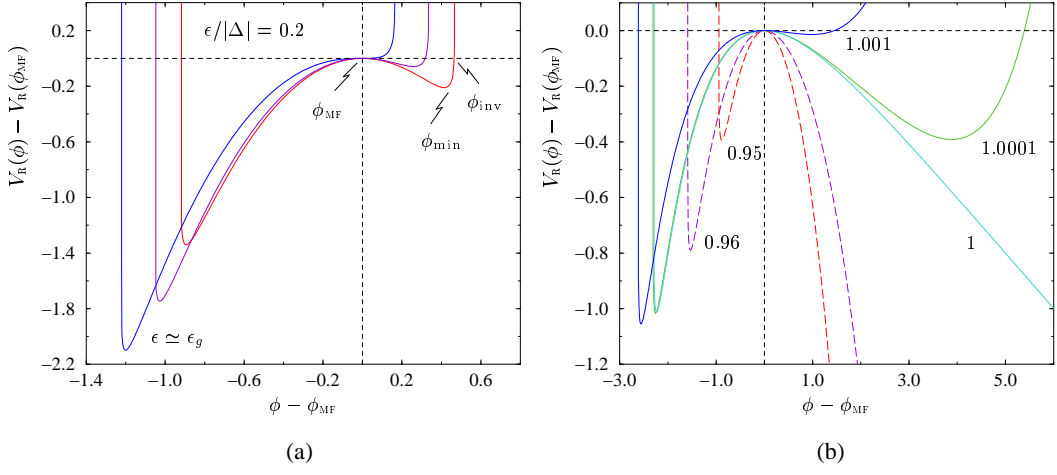}
\end{center}
\caption{\small
        Rescaled potential $V_{\smc{r}} (\phi) - V_{\smc{r}}
        (\phi_{\smc{mf}})$ versus the variable $\phi' = \phi -
        \phi_{\smc{mf}}$ for $\alpha = 0.25$, $\zeta = 0.05$; (a)
        $\epsilon / |\Delta| = 0.2$, $\epsilon / |\Delta| = 0.3 <
        \epsilon_g / |\Delta|$ and $\epsilon / |\Delta| = 0.37622
        \simeq \epsilon_g / |\Delta|$. The inversion point is
        indicated with $\phi_{\text{inv}}$, while the minimum of the
        potential is indicated with $\phi_{\text{min}}$;
        (b) $\epsilon / |\Delta| = 0.95 \gtrsim \Phi_2$, $\epsilon /
        |\Delta| = 0.96$, $\epsilon / |\Delta| = 1$, $\epsilon /
        |\Delta| = 1.0001$ and $\epsilon / |\Delta| = 1.001 \lesssim
        \Phi_3$.
        }
\label{fig:pote1}
\end{figure}

Typical shapes of the potential for a fixed values of the scattering
amplitude ($\alpha = 0.25$) and the magnetic impurity concentration
($\zeta = 0.05$) and in two different regions of the energy are shown
in Fig.~\ref{fig:pote1}. A bounce solution with minimum action exists
in the regions $\epsilon < \epsilon_1 = \epsilon_g$
(Fig.~\ref{fig:pote1} (a)) and $\epsilon_2 < \epsilon < \epsilon_3$
(Fig.~\ref{fig:pote1} (b)), where $\Phi_l \equiv \epsilon_l/|\Delta|$,
while outside both the unique solution is the homogeneous one, $\phi =
\phi_{\smc{mf}}$. In fact when the energy approaches one of the hard
edges predicted by the mean-field theory, $\epsilon \to \epsilon_1^-$, 
$\epsilon \to \epsilon_2^+$ or $\epsilon\to \epsilon_3^-$, the maximum
of the potential for $\phi = \phi_{\smc{mf}}$ merges with the minimum
of the potential corresponding to the lowest value of the action.

Now, for simplicity, let us first focus on the quasi-one dimensional
case. Later we will generalise the discussion to the $d$-dimensional
case. The symmetry broken solution~\eqref{eq:bounc} involves the real
instanton action
\begin{align}
  S_{\text{inst}} &= 4 \pi \nu |\Delta| \xi S_{\phi} &
  S_{\phi} &= \int_{\phi_{\smc{mf}}}^{\phi_{\text{inv}}} d \phi
  \sqrt{V_{\smc{r}} (\phi_{\smc{mf}}) - V_{\smc{r}} (\phi)} \; ,
\label{eq:insta}
\end{align}
where $\phi_{\text{inv}}$ represents the `classical turning point',
$V_{\smc{r}} (\phi_{\text{inv}}) = V_{\smc{r}} (\phi_{\smc{mf}})$. As
the energy approaches one of the three edges ($\epsilon \to
\epsilon_1^-$, $\epsilon \to \epsilon_2^+$ or $\epsilon\to
\epsilon_3^-$), the minimum of the potential~\eqref{eq:fpote}
corresponding to the lowest value of the instanton action
\eqref{eq:insta} disappears, merging with the maximum at the
mean-field point $\phi_{\smc{mf}}$. Developing the function $g (u ,
\zeta , \alpha)$ \eqref{eq:gufun} around one of the extrema
$\epsilon_l / |\Delta|$,
\begin{displaymath}
  u - u_0^{l} = (-1)^{l} \sqrt{\Frac{2}{|g'' (u_0^{l} , \zeta ,
  \alpha)|}} \left[ (-1)^{l} \left(\Frac{\epsilon -
  \epsilon_l}{|\Delta|} \right) \right]^{1/2} \; ,
\end{displaymath}
it is possible to deduce the expansion of the potential $V_{\smc{r}}
(\phi)$ in powers of $(\phi - \phi_{\smc{mf}})$ around each edge:
\begin{equation}
  V_{\smc{r}} (\phi) - V_{\smc{r}} (\phi_{\smc{mf}}) \simeq - A_l
  \left[(-1)^{l} \left(\Frac{\epsilon - \epsilon_l}{|\Delta|}\right)
  \right]^{1/2} (\phi - \phi_{\smc{mf}})^2 - (-1)^{l} B_l (\phi -
  \phi_{\smc{mf}})^3 \; .
\label{eq:pdeve}
\end{equation}
Here the positive dimensionless coefficients $A_l (u_0^{l}, \zeta ,
\alpha)$ and $B_l (u_0^{l}, \zeta , \alpha)$ depend solely on the
scattering amplitude $\alpha$ and the magnetic impurity concentration
$\zeta$.

As the energy approaches one of the hard edges, from below or above
depending on whether we are dealing with $\Phi_1$ and $\Phi_3$ or $\Phi_2$,
the expansion~\eqref{eq:pdeve} permits one to obtain an analytic solution
for $S_\phi$,
\begin{equation*}
  S_{\phi} = \Frac{4}{15} \Frac{A_l^{5/2}}{B_l^2} \left[(-1)^{l}
  \left(\Frac{\epsilon - \epsilon_l}{|\Delta|}\right)\right]^{5/4}
  \; ,
\end{equation*}
and the bounce solution
\begin{displaymath}
  \phi(x) - \phi_{\smc{mf}} = \Frac{A_l}{B_l}\Frac{1}{\cosh^2 x/2r_0}
  \; ,
\end{displaymath}
where the extent of the instanton is set by
\begin{equation*}
  r_0 (\epsilon) = \Frac{\xi}{A_l^{1/2}} \left[(-1)^{l}
  \left(\Frac{|\Delta|}{\epsilon - \epsilon_l}\right)\right]^{1/4} \;
  .
\end{equation*}
Therefore, on approaching the edge from the `gapped region'
(i.e. $\epsilon \to \epsilon_{1,3}^{-}$ or $\epsilon \to
\epsilon_{2}^{+}$) the size of the `droplet' $r_0 (\epsilon)$ goes to
infinity. Moreover, the coefficient $A_1$ goes to zero at the gapless
point and analogously $A_2$ and $A_3$ go to zero when the impurity
band merges with the continuum.

When in the vicinity of the mean-field gap edges, a generalisation of 
the quasi one-dimensional results to higher dimensions ($1<d<6$) can be
developed by dimensional analysis. Using the approximation~\eqref{eq:pdeve}, 
the bounce configuration is shown to have the scaling form
\begin{displaymath}
  \phi(\vect{r}) -\phi_{\smc{mf}} = \left[\Frac{\xi}{r_0
  (\epsilon)}\right]^2 \Frac{1}{B_l} f(\vect{r}/ r_0 (\epsilon)) \; .
\end{displaymath}
Substituting this relation in the expression of the instanton action,
one finds that:
\begin{equation}
  S_{\text{inst}} = 4 a_d \pi g \left(\Frac{\xi}{L}\right)^{d-2}
  B_l^{-2} A_l^{(6-d)/2} \left[(-1)^{l} \left(\Frac{\epsilon -
  \epsilon_l}{|\Delta|}\right)\right]^{(6-d)/4} \; ,
\label{eq:dgenu}
\end{equation}
where $g=\nu D L^{d-2}$ is the bare dimensionless conductance and
  $a_d$ is a numerical constant\footnote{\ 
  Specifically, $a_d = \int d \vect{u} [(\nabla_\vect{u} f)^2 + f^2
  (\vect{u}) - f^3(\vect{u})]$.}.
This closes our discussion of the particular saddle-point solution 
together with the corresponding action. However, to complete the analysis
it is necessary to explore the influence of fluctuations around the
instanton solution.

\begin{figure}
\begin{center}
\includegraphics[width=0.6\linewidth,angle=0]{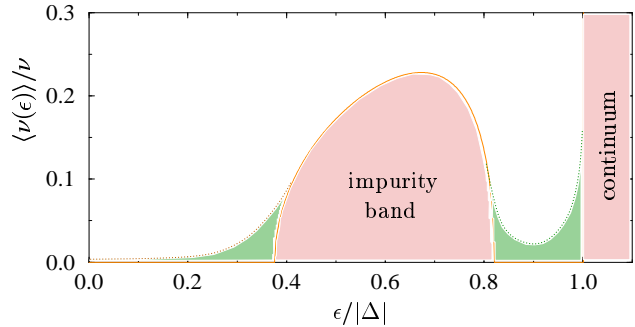}
\end{center}
\caption{\small
        Smearing of the gap edges due to optimal configurations of the
        random impurity potentials. The three edges, $\epsilon_g$,
        $\epsilon_2$ and $\epsilon_2$  become the mobility edges
        separating the regions of bulk delocalised states from the
        localised tail states.
        }
\label{fig:mobil}
\end{figure}

\subsection{Fluctuations}
\label{sec:fluct}
Here we only sketch the important aspects of the fluctuation analysis, 
referring to Ref.~\cite{lamacraft_simons} for a more detailed discussion
in the context of the Born scattering limit. Generally, field fluctuations 
around the instanton solution can be separated into `radial' and `angular' 
contributions. The former involve fluctuations of the diagonal elements 
$\hat{\theta}$, while the latter describe rotations including those 
Grassmann transformations which mix the $\smc{bf}$ sector. Both classes 
of fluctuations play an important role. 

Dealing first with the angular fluctuations, supersymmetry breaking of
the bounce is accompanied by the appearance of a Grassmann zero mode
separated by an energy gap from higher excitations. This Goldstone
mode restores the global supersymmetry of the theory. Crucially, this
mode ensures that the saddle-point respects the normalisation
condition $\langle \mathcal{Z}[0] \rangle_{W,V} =1$. Associated with
radial fluctuations around the bounce, there exists a zero mode due to
translational invariance of the solution, and a negative energy mode
(c.f. Ref.~\cite{coleman}). 

Combining these contributions, one obtains the following expression for 
the local complex DoS in the tail regions ($\epsilon \lesssim
\epsilon_g$, $\epsilon \gtrsim \epsilon_2$ and $\epsilon \lesssim
\epsilon_3$),
\begin{equation*}
  \langle \nu (\epsilon) \rangle_{V,S} \Simi_{\epsilon \simeq
  \epsilon_l} \nu \int d\vect{r}\left[\sinh \phi (\vect{r}) -
  \sinh \phi_{\smc{mf}}\right] |\chi_0 (\vect{r})|^2
  \sqrt{\Frac{LS_{\phi}}{\xi}} \; e^{- S_{\text{inst}}} \; , 
\end{equation*}
where $\chi_0 (\vect{r})$ represents the eigenfunction for the Grassmann zero
mode, $\sqrt{LS_{\phi} / \xi}$ is the Jacobian associated with the
introduction of the collective coordinate~\cite{coleman} and
$S_{\text{inst}}$ denotes the instanton action~\eqref{eq:dgenu}. 
Thus, to exponential accuracy, the complex local DoS in the tail
region becomes non-zero only in the vicinity of the bounce. 

\section{Discussion}
\label{sec:discu}

This concludes our investigation of the role of quenched classical magnetic 
impurities on the weakly disordered superconducting system. The results 
above show that the gap structure predicted by the mean-field theory for 
magnetic impurities in both the unitarity and Born scattering limit is 
untenable. The hard gap edge(s) predicted by the mean-field theory are 
softened by the nucleation of domains or `droplets' of localised tail 
states. The mean-field gap edges become mobility edges separating bulk 
delocalised quasi-particle states from localised tail states (see 
Fig.~\ref{fig:mobil}). The latter are induced by optimal fluctuations of 
the non-magnetic random impurity potential which increase the effective
spin scattering rate. 

How significant are these results for experiment? The nucleation of
sub-gap states will lead to the softening of the transition from the
metallic to the superconducting phase which would be revealed in
measurements of the heat capacity close to the bulk $T_c$. Similarly,
since the tail state regions are broad on the scale of the coherence
length (and broaden close to the quasi-particle energy gap), one can
expect the existence of sub-gap states to be revealed in measurements
of the tunneling density of states close to the mean-field
gap. However, it should be noted that the strength of arising
from the sub-gap states will compete with the weight arising from
dynamical spin fluctuations. The latter will typically give rise to
power law tails in the sub-gap density of
states~\cite{fb_spin_dyn}.

Finally, our considerations of the density of states focussed largely on 
energy scales $\epsilon \sim \epsilon_g$. Here the generating function is 
dominated by the homogeneous mean-field and bounce configurations of the 
non-linear $\sigma$-model action. However, within the gapless phase (i.e. 
as $\epsilon \to 0$), field fluctuations $Q$ which commute with both 
$\sigma_1^{\smc{ph}}$ and $\sigma_3^{\smc{ph}}$ become massless. These 
fluctuations, which are parameterised by transformations $Q=TQ_{\smc{mf}}
T^{-1}$ where $T=\mathbb{I}_{\smc{ph}}\otimes\mathbb{I}_{\smc{sp}}\otimes 
t$ and $t=\gamma(t^{-1})^T\gamma^{-1}$, are controlled by a non-linear 
$\sigma$-model defined on the manifold $T\in {\rm OSp}(2|2)/{\rm GL}(1|1)$ 
(symmetry class D in the classification of Ref.~\cite{zirnbauer}). The
latter reflect quantum interference effects in the particle/hole channel 
and substantially modify the low-energy, long-range spectral and transport 
properties of both the bulk system and the low-energy states inside a 
localised domain. For a comprehensive discussion of their effect, we refer
to the comprehensive discussions in
Refs.~\cite{bundschuh,senthil_fisher,read_green,bocquet_serban,lamacraft_simons}.

\paragraph{Acknowledgments:} We are grateful to Austen Lamacraft and
Robert Moir for valuable discussions.

\end{document}